\documentclass[aps,preprint,groupedaddress,showpacs]{revtex4}

\usepackage{graphicx}
\usepackage{dcolumn}
\usepackage{bm}

\newcommand{\pt}{\mbox{$p_T$} }

\newcommand{\gevc}{\mbox{${\rm GeV}/c$} }
\newcommand{\gevcc}{\mbox{${\rm GeV}/c^2$} }

\newcommand{\mm}{\mbox{$\mu{\rm m}$} }

\newcommand{\Bm}{$B$ meson }
\newcommand{\Bms}{$B$ mesons }

\newcommand{\intlumx}{98 $\pm$ 4 pb$^{-1}$ }
\newcommand{\bdecay}{$B^{\pm} \to J/\psi \, K^{\pm}$ }

\newcommand{\jpsi}{$J/\psi$ }

\newcommand{\bspace}{\hspace*{-0.065in}}

\begin{document}



\title{Measurement of the $B^+$ Total Cross Section and $B^+$ Differential 
Cross Section $d\sigma/dp_T$ in $p \bar p $
Collisions at $\sqrt{s}=1.8$ TeV}



\date{\today}
\maketitle
\font\eightit=cmti8
\def\r#1{\ignorespaces $^{#1}$}
\hfilneg
\begin{sloppypar}
\noindent
D.~Acosta,\r {13} T.~Affolder,\r {24} H.~Akimoto,\r {46}
M.~G.~Albrow,\r {12} P.~Amaral,\r 9 D.~Ambrose,\r {33}   
D.~Amidei,\r {26} K.~Anikeev,\r {25} J.~Antos,\r 1 
G.~Apollinari,\r {12}
% ...
\end{sloppypar}
\vskip .026in
\begin{center}
(CDF Collaboration)
\end{center}

\vskip .026in
\begin{center}
\r 1  {\eightit Institute of Physics, Academia Sinica, Taipei, Taiwan 11529, 
Republic of China} \\
\r 2  {\eightit Argonne National Laboratory, Argonne, Illinois 60439} \\
\r 3  {\eightit Istituto Nazionale di Fisica Nucleare, University of Bologna,
I-40127 Bologna, Italy} \\
\r 4  {\eightit Brandeis University, Waltham, Massachusetts 02254} \\
\r 5  {\eightit University of California at Davis, Davis, California  95616} \\
\r 6  {\eightit University of California at Los Angeles, Los 
Angeles, California  90024} \\  
\r 7  {\eightit Instituto de Fisica de Cantabria, CSIC-University of Cantabria, 
39005 Santander, Spain} \\
\r 8  {\eightit Carnegie Mellon University, Pittsburgh, PA  15218} \\
\r 9  {\eightit Enrico Fermi Institute, University of Chicago, Chicago, 
Illinois 60637} \\
\r {10}  {\eightit Joint Institute for Nuclear Research, RU-141980 Dubna, Russia}
\\
\r {11} {\eightit Duke University, Durham, North Carolina  27708} \\
\r {12} {\eightit Fermi National Accelerator Laboratory, Batavia, Illinois 
60510} \\
\r {13} {\eightit University of Florida, Gainesville, Florida  32611} \\
\r {14} {\eightit Laboratori Nazionali di Frascati, Istituto Nazionale di Fisica
               Nucleare, I-00044 Frascati, Italy} \\
\r {15} {\eightit University of Geneva, CH-1211 Geneva 4, Switzerland} \\
\r {16} {\eightit Glasgow University, Glasgow G12 8QQ, United Kingdom}\\
\r {17} {\eightit Harvard University, Cambridge, Massachusetts 02138} \\
\r {18} {\eightit Hiroshima University, Higashi-Hiroshima 724, Japan} \\
\r {19} {\eightit University of Illinois, Urbana, Illinois 61801} \\
\r {20} {\eightit The Johns Hopkins University, Baltimore, Maryland 21218} \\
\r {21} {\eightit Institut f\"{u}r Experimentelle Kernphysik, 
Universit\"{a}t Karlsruhe, 76128 Karlsruhe, Germany} \\
\r {22} {\eightit Center for High Energy Physics: Kyungpook National
University, Taegu 702-701; Seoul National University, Seoul 151-742; and
SungKyunKwan University, Suwon 440-746; Korea} \\
\r {23} {\eightit High Energy Accelerator Research Organization (KEK), Tsukuba, 
Ibaraki 305, Japan} \\
\r {24} {\eightit Ernest Orlando Lawrence Berkeley National Laboratory, 
Berkeley, California 94720} \\
\r {25} {\eightit Massachusetts Institute of Technology, Cambridge,
Massachusetts  02139} \\   
\r {26} {\eightit University of Michigan, Ann Arbor, Michigan 48109} \\
\r {27} {\eightit Michigan State University, East Lansing, Michigan  48824} \\
\r {28} {\eightit University of New Mexico, Albuquerque, New Mexico 87131} \\
\r {29} {\eightit The Ohio State University, Columbus, Ohio  43210} \\
\r {30} {\eightit Osaka City University, Osaka 588, Japan} \\
\r {31} {\eightit University of Oxford, Oxford OX1 3RH, United Kingdom} \\
\r {32} {\eightit Universita di Padova, Istituto Nazionale di Fisica 
          Nucleare, Sezione di Padova, I-35131 Padova, Italy} \\
\r {33} {\eightit University of Pennsylvania, Philadelphia, 
        Pennsylvania 19104} \\   
\r {34} {\eightit Istituto Nazionale di Fisica Nucleare, University and Scuola
               Normale Superiore of Pisa, I-56100 Pisa, Italy} \\
\r {35} {\eightit University of Pittsburgh, Pittsburgh, Pennsylvania 15260} \\
\r {36} {\eightit Purdue University, West Lafayette, Indiana 47907} \\
\r {37} {\eightit University of Rochester, Rochester, New York 14627} \\
\r {38} {\eightit Rockefeller University, New York, New York 10021} \\
\r {39} {\eightit Rutgers University, Piscataway, New Jersey 08855} \\
\r {40} {\eightit Texas A\&M University, College Station, Texas 77843} \\
\r {41} {\eightit Texas Tech University, Lubbock, Texas 79409} \\
\r {42} {\eightit Institute of Particle Physics, University of Toronto, Toronto
M5S 1A7, Canada} \\
\r {43} {\eightit Istituto Nazionale di Fisica Nucleare, University of Trieste/
Udine, Italy} \\
\r {44} {\eightit University of Tsukuba, Tsukuba, Ibaraki 305, Japan} \\
\r {45} {\eightit Tufts University, Medford, Massachusetts 02155} \\
\r {46} {\eightit Waseda University, Tokyo 169, Japan} \\
\r {47} {\eightit University of Wisconsin, Madison, Wisconsin 53706} \\
\r {48} {\eightit Yale University, New Haven, Connecticut 06520} \\
\r {(\ast)} {\eightit Now at Northwestern University, Evanston, Illinois 
60208} \\
\r {(\ast\ast)} {\eightit Now at University of California, Santa Barbara, CA
93106}
\end{center}


\begin{center}
{{\bf Abstract}\\}
\end{center}
We present measurements of the $B^+$ meson total cross section and 
  differential cross section $d\sigma/ dp_T$. The measurements use a \intlumx sample
  of $p \bar p$ collisions at $\sqrt{s}=1.8$ TeV collected by the CDF
  detector.  Charged $B$ meson candidates are reconstructed through the
  decay $B^{\pm} \rightarrow J/\psi K^{\pm}$ with $J/\psi\rightarrow \mu^+
  \mu^-$. The total cross section, measured in the central rapidity region
  $|y|<1.0$ for $p_T(B)>6.0$ GeV/$c$, is $3.6 \pm 0.6 ({\rm stat} \oplus
  {\rm syst)} \mu$b. 
  The measured differential cross section is substantially larger than   
  typical QCD predictions calculated to next-to-leading order.





%
\newpage

\section{Introduction \label{s:introduction} }

Quantum Chromodynamics (QCD) can be used to compute the expected cross
sections for the production of heavy quarks at hadron collider
energies. Calculations of the hard-scattering cross section have been
carried out to next-to-leading order in perturbation
theory~\cite{QCDcalc}.  Experimental measurements must show that 
these predictions
provide an adequate description of the cross section at 1.8 TeV before
they can be confidently extrapolated to higher energies or more exotic
phenomena. Unfortunately the QCD predictions are affected by large
theoretical uncertainties such as the dependence on the choice of the
factorization and renormalization scales, the parton density
parameterization and the $b$ quark mass~\cite{Mangano}.

Experiments at CERN~\cite{ua1} and at the Tevatron~\cite{tevatron}
have shown that the $b$ quark production cross section is higher than
the theoretical predictions obtained with the standard choice of
parameters by about a factor of 2--3. Closer agreement between theory
and the experimental measurements can be achieved by choosing rather
extreme values of the theoretical parameters~\cite{Mangano}.
It has also been suggested that the large discrepancy could
be explained by pair production of light gluinos that decay
into bottom quarks and bottom squarks~\cite{Berger}.

This paper describes a measurement of the $B^+$ meson total cross section 
and differential
cross section $d\sigma/dp_T$ in hadronic collisions using fully reconstructed 
$B^{\pm}$ mesons decaying into the exclusive
final state $J/\psi\,K^{\pm}$. The measurement uses a data sample of
\intlumx collected by the Collider Detector
at Fermilab (CDF) experiment from $p \bar p$ collisions
with a center-of-mass energy of 1.8 TeV produced by the Fermilab
Tevatron.  The data were collected in the run period from 1992 to 1995
which is referred to as Run 1. Our previously published 
result~\cite{CDFold} based upon $19.3 \pm 0.7 $ pb$^{-1}$ 
of data (Run 1A) found that the total
cross section for $p_T(B) > 6.0 $ \gevc and $|y|<1.0$
is $\sigma_{B}$~=~2.39~$\pm$ 0.54(stat $\oplus$ syst) $\mu$b.

The paper is organized as follows. In Section~\ref{s:previous} we review previous 
measurements of the $B$ cross section using exclusive $B$ decays. In 
Section~\ref{s:detector} we briefly describe the components of
the CDF detector 
relevant to the analysis presented in this paper. The data collection, event selection procedures  
and the reconstruction of 
$B^{\pm} \rightarrow J/\psi K^{\pm}$ are discussed in Section ~\ref{s:sample}. The 
measurement 
of the differential and total cross sections is presented in 
Sections~\ref{s:measurement} and~\ref{s:total}, respectively. 

\section {Previous Measurement of the $B$ Production 
Cross Section \label{s:previous}}
  

The Run 1A measurement of the $B$
meson differential cross section was determined from fully reconstructing the 
decays $B^{\pm} \rightarrow J/\psi K^{\pm}$ and $B^0 \rightarrow J/\psi
K^{*0}(892)$~\cite{CDFold}.  The measurement of the $B$ transverse
momentum spectrum showed that next-to-leading-order QCD adequately
described the shape of this distribution for $p_T > 6.0$ GeV/$c$. 
In the Run 1A
publication, CDF used a branching ratio $BR(B^+\rightarrow J/\psi K^+)$~=~$(11.0 \pm
1.7)\times 10^{-4}$  
and a product of branching fractions $\mathcal {B}
= BR(B^+\rightarrow J/\psi K^+) \times BR(J/\psi
\rightarrow \mu^+ \mu^-)$~=~$(6.55 \pm 1.01)\times 10^{-5}$
~\cite{Cleo}.  
The current world average for $BR(B^+\rightarrow
J/\psi K^+)$ is $(10.0 \pm 1.0)\times 10^{-4}$  
which yields $\mathcal{B}=(5.88 \pm 0.60)\times 10^{-5}$.  
The change in the branching
fractions scales the published result up by about 10\% to
$\sigma_{B}$($p_{T}$ $>$ 6.0 GeV/$c$, $|y|<1.0$) =
$2.66 \pm 0.61$ (stat $\oplus$ syst)$\mu$b.

This paper updates the measurement presented in 1995 by using the
complete Run 1 data sample of \intlumx\bspace.
For this measurement, we use 
only the decay mode $B^{\pm} \rightarrow J/\psi
K^{\pm}$ where we require both muon candidates from the $J/\psi$ decay
to be well measured by the silicon vertex detector (SVX).  Such a
restriction allows us to use fewer selection requirements since the
decay mode $B^{\pm} \rightarrow J/\psi K^{\pm}$ has a lower combinatorial
background than $B^0 \rightarrow J/\psi K^{*0}$, and the SVX information
enables us to substantially reduce the prompt background.  Moreover,
several of the efficiencies are measured using a large sample of
$J/\psi \rightarrow \mu^+ \mu^-$ candidates rather
than relying on Monte Carlo calculations for detailed modeling of detector 
effects.

\section{The CDF detector\label{s:detector}} 

The CDF detector is described in detail in~\cite{CDFdet}.  We
summarize here the features of the detector subsystems that are
important for this analysis.  The CDF coordinate system has the $z$
axis pointing along the proton beam momentum, and the angle $\phi$ is
measured from the plane of the Tevatron storage ring.  The transverse
($r$-$\phi$) plane is normal to the proton beam.

The CDF experiment uses three separate detectors for tracking
charged particles: the silicon vertex detector (SVX), the vertex
detector (VTX), and the central tracking chamber (CTC).  These devices
are immersed in a magnetic field of 1.4 Tesla pointed along the $-z$
axis generated by a superconducting solenoid of length 4.8 m and
radius 1.5 m.

The innermost device is the SVX~\cite{CDFsvx} which provides spatial
measurements in the $r$-$\phi$ plane.  The SVX consists of two
cylindrical barrels that cover a region 51 cm long in $z$.
Each barrel consists of four layers of
silicon strip sensors with strips oriented parallel to the beam axis.
The distribution of the $p \bar p$ collisions along the beamline is Gaussian 
in $z$ with a $\sigma$ of about 30 cm. Therefore only about $60$\% of
all $J/\psi\rightarrow \mu^+\mu^-$ events have both muon tracks
reconstructed in the SVX.  

The SVX is surrounded by the VTX, a set of time projection chambers
which measure the $z$ coordinate of the $ p \bar p$ interaction
(primary vertex). Surrounding the SVX and the VTX is the CTC. The CTC
is a 3.2 m long cylindrical drift chamber with 84 layers of sense
wires ranging in radius from 31 cm to 133 cm.  The combined momentum
resolution of the tracking chambers is $\delta p_T/p_T$~=~$[(0.0009 ~p_T)^2 
+ (0.0066)^2]^{1/2}$ where $p_T$ is the component of
the momentum transverse to the $z$ axis and is measured in GeV/$c$.
Charged track trajectories reconstructed in the CTC that are matched
to strip clusters in the SVX have an impact parameter resolution of
$\sigma_d(p_T)=(13+40/p_T)$ $\mu$m~\cite{resol} with $p_T$ in units
of GeV/$c$.  The track impact parameter $d$ is defined as the distance
of closest approach of the track helix to the beam axis measured in
the plane perpendicular to the beam.

The central muon system consists of three components (CMU, CMP and
CMX) and detects muons with $p_T \ge 1.4$ GeV/$c$ in the
pseudorapidity range $|\eta|<1.0$. The CMU system covers the region
$|\eta|<0.6$ and consists of four layers of drift chambers
outside the hadron calorimeter.
Outside the CMU there is an
additional absorber of 60 cm of steel followed by four layers of drift
chambers (CMP). The CMX system extends the coverage to pseudorapidity
$0.6<|\eta|<1.0$ but is not used in this analysis.

CDF employs a three level trigger system.  The first two levels are
implemented in custom electronics.  To select events in the third
level, we employ a CPU farm using
a version of the CDF event reconstruction program optimized for speed.

\section{Data Sample Selection\label{s:sample}} 


%
%

\subsection{Dimuon Trigger\label{s:trigger}}

The data sample consists of events that pass the 
$J/\psi\rightarrow \mu^+ \mu^-$ trigger.
In the first level of this trigger, we require two muon track segments 
in the central muon chambers separated by at
least $5^\circ$ in azimuth. The trigger efficiency 
for each muon at Level 1 rises from 50\% for
$p_T$ = 1.7~GeV$/c$ to 95\% for $p_T$ = 3.3~GeV$/c$.  

In the second level, we require muon segments found 
in Level~1 to be associated
with tracks identified by the Central Fast Tracker (CFT) \cite{CFT}.
The resolution of the CFT is $\delta p_T/p_T^2\approx 0.03\,({\rm
GeV}/c)^{-1}$.  In Run 1A and for a subset of the Run 1B data, we
required one of the two muons to be matched to a CFT track with $p_T$
greater than about 3~GeV$/c$ while in the bulk of the Run~1B sample, 
we required two muon segments to have an associated track with a
threshold of about 2~GeV$/c$.  In Run 1A (1B), the extrapolation of
the track was required to be typically within 10$^\circ$ (5$^\circ$)
of the muon segment.  The efficiency of the track requirements was
measured in a $J/\psi$ data sample using events in which the muon under study
need not have satisfied the requirements for the event to be accepted.  The
efficiency for the nominal 2 (3) GeV$/c$ threshold rose from 50\% of
the plateau efficiency at 1.95 (3.05) GeV$/c$ to 95\% of the plateau
efficiency at 2.2 (3.4) GeV$/c$.  That plateau efficiency changed over
the course of the run because of aging of the CTC and subsequent
modifications to the CFT algorithms.  That dependence on time is
accounted for in the calculation of the trigger efficiencies.

The Level 3 software trigger required two muon candidates with an 
effective mass in the $J/\psi$ mass region after full 
reconstruction. Runs with known hardware problems for muons were removed 
yielding for this analysis a total Run 1 luminosity of 98 pb$^{-1}$.

\subsection{$J/\psi$ Reconstruction\label{s:psireco}}

Background events in the dimuon sample collected with these triggers
are suppressed by applying additional muon selection cuts. Track
quality requirements are used to reduce the backgrounds arising from
poor track measurements. Tighter cuts are imposed
on the correlation between the track in the muon chamber and the
extrapolated CTC track. 

The transverse momentum of each muon from the $J/\psi$ for Run 1A is
required to be greater than 1.8 GeV/$c$ with one muon of the pair
greater than 2.8 GeV/$c$. For Run 1B, both muons are required to have
a transverse momentum greater than 2.0 GeV/$c$. Events passing both
the trigger and $p_{T}$ requirements identical to those of Run 1A are also
accepted. The muons must have opposite charge and the separation in
$z$ between the two tracks must be less than 5.0 cm at the point of
closest approach to the beamline.  The $z$ coordinate of the decay
vertex is required to be within $\pm 60 $ cm of the detector center.

The invariant mass and uncertainty ($\sigma_m$) of the $J/\psi$ candidates 
are calculated after constraining the two
muon tracks to come from a common point in space (vertex constraint)
to improve the mass resolution. The width of the reconstructed
$J/\psi$ mass peak is 16 MeV/$c^2$. The signal region is defined to be
those dimuon candidates with reconstructed mass within 3.3$\sigma_m$
of the known $J/\psi $ mass~\cite{PDG}. We find
$(8.7 \pm 0.2)\times 10^4$ $J/\psi$ over background. In this analysis, 
the two muons from the $J/\psi $ decay are required to be 
reconstructed in the silicon detector. 

\subsection{Primary Vertex Selection\label{s:primary}}

Knowledge of the distance between the primary $p \bar p$ interaction
vertex and the secondary decay vertex in the transverse plane is
crucial to this analysis since the $B$ meson proper lifetime is used
to discriminate between $B$ mesons and background events. We find the 
transverse position of the primary vertex using the
average beamline calculated for each Tevatron $p \bar p$ store~\cite {CDFlife}.
The longitudinal coordinate of the primary vertex $(z)$ is measured
using data from the VTX detector. The slopes and intercepts of the
run-averaged beam position are combined with the event-by-event $z$
locations of the vertices to determine the vertex position. The primary
vertex uncertainties $\sigma_x$, $\sigma_y$ and $\sigma_z$ are estimated to
be 25, 25 and 300 $\mu$m, respectively.

\subsection{$B$ Reconstruction\label{s:breco}}

To select charged $B$ candidates we considered each charged particle
track as a kaon candidate to be combined with a $J/\psi$. A charged
track in an event is combined with the two muons if the $z_{0}$
parameter of the track is within 5 cm of the $z$ position of the
$J/\psi$ candidate decay vertex.  The exit radius of the kaon
candidate, which corresponds to the radius at which the track
trajectory intersects the plane of the CTC endplate, is required to be
greater than $110$ cm to limit the search to a region of high tracking
efficiency. A cut on the kaon transverse momentum of $p_{T}>1.25$
GeV/$c$ is imposed to reduce the large combinatorial background.  This
cut is very effective since kaons from $B$ meson decay have a
considerably harder $p_{T}$ spectrum than particles from the
underlying event and from events with prompt $J/\psi$ production. The
muon and kaon tracks are constrained to come from a common point of
origin and the mass of the $\mu^+$ $\mu^-$ pair is constrained to the
known $J/\psi$ mass.  Since the intrinsic width of the $J/\psi$ is
significantly smaller than our experimental resolution, the mass
constraint improves the resolution of the reconstructed $B$ mass.
 
The $p_{T}$ of each $B$ candidate is required to be greater than
$6.0$ GeV/$ c$. The proper decay length is required to be greater than
100 $\mu$m to suppress backgrounds associated with prompt $J/\psi$
mesons. The signed proper decay length in the $B$ rest frame is
defined as
\begin{equation}
  ct(B) = \frac{\overrightarrow{X_{J/\psi}} \cdot
  \overrightarrow{p^{B}_{T}}}{p_{T}^{B}} \cdot \frac{1}{(\beta
  \gamma)^{B}} = \frac{M_{B} \overrightarrow{X_{J/\psi }} \cdot
  \overrightarrow{p^{B}_{T}}}{(p^{B}_{T})^{2}}
\label{eq:ctb-=-frac}
\end{equation}
where
\begin{equation}
  \overrightarrow{X_{J/\psi }} = (x_{J/\psi }-x_{PV}) \hat{i} +
  (y_{J/\psi }-y_{PV}) \hat{j}
\end{equation}
and $(\beta \gamma)^{B}$ is the relativistic boost of the \Bm\bspace.
The $(x_{J/\psi},y_{J/\psi})$ are the transverse coordinates of the
\jpsi decay vertex, and the $(x_{PV},y_{PV})$ are the transverse
coordinates of the event primary vertex.  
The intersection of the muon tracks as measured in the SVX determines 
the location of the $B$ meson decay. 

\begin{figure}
\centering
\includegraphics[width=10cm]{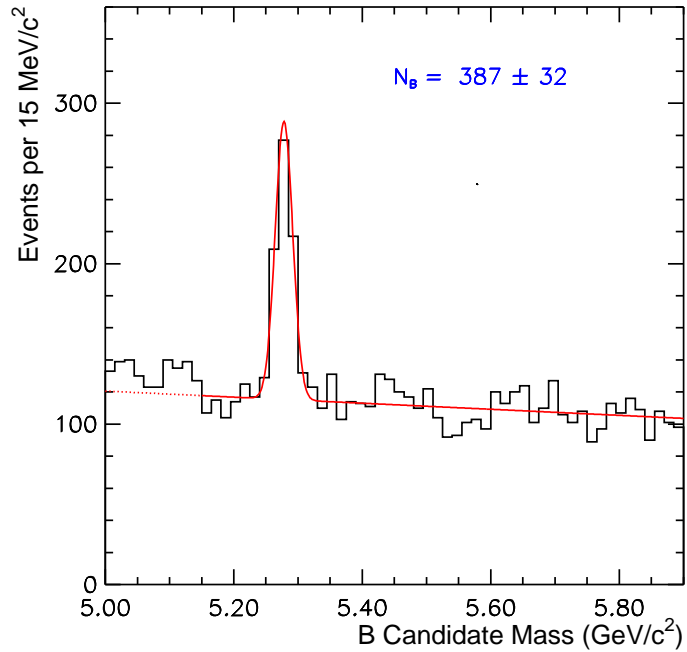}
\caption{$B^{\pm}$ invariant mass distribution reconstructed from the decay \bdecay\bspace.
The curve is a binned fit to a Gaussian distribution plus linear 
background and 
is for illustration only.}
\label{fig:BmassDist}
\end{figure}

The $B^{\pm}$ candidate mass distribution is shown in Figure
\ref{fig:BmassDist}. The distribution is fit with a Gaussian signal
function plus a linear background using an unbinned maximum likelihood
fit. The region below 5.15 GeV/$c^2$ has been excluded from the fit
since it includes contributions from partially reconstructed
higher-multiplicity $B$-decay modes. The fit yields $387 \pm 32$
$B^{\pm}$ mesons.

\section{Differential cross section\label{s:measurement}} 

To measure the differential cross section, we divide the $B$ candidate sample
into four \pt ranges. The invariant mass distributions for each of the
\pt ranges are then fitted using an unbinned maximum likelihood fit
which is described in Section A. The determination of the geometric
acceptance, the efficiencies and the luminosity are described in Sections B, C
and D respectively.  The systematic uncertainties are discussed in Section E, and
the results are presented in Section F.

\subsection{Fitting Technique}

\begin{figure}[htbp]
\centering
\includegraphics[width=12cm]{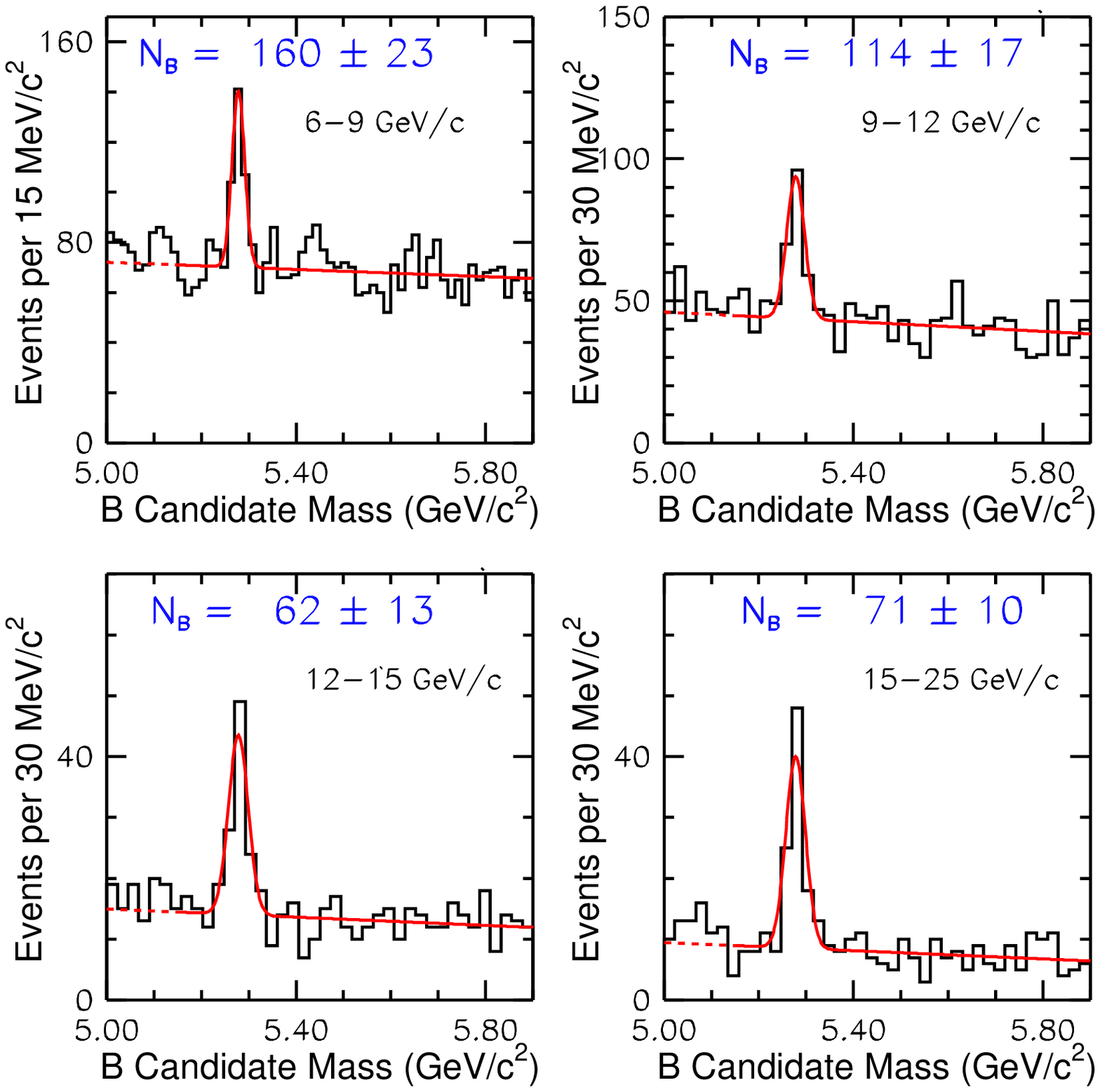}
\caption{$B^{\pm}$ candidate mass distribution for the four \pt ranges.
The curve is a binned fit to a Gaussian distribution plus linear 
background and 
is for illustration only.}
\label{fig:Bmass4pt}
\end{figure}

To measure the $B^+$ meson differential cross section as a function of
\pt\bspace, the $B$ candidate sample is divided into four \pt
bins: 6--9, 9--12, 12--15, and 15--25 \gevc\bspace.  The invariant
mass distribution for each of the \pt ranges is then fitted using an
unbinned maximum likelihood fit to determine the number of $B$
candidates in each $p_T$ range, as shown in Figure \ref{fig:Bmass4pt}. The
likelihood function is a Gaussian signal plus a linear background:
\begin{equation}
  {\mathcal{L}} =  \frac{N_{sig}}{N_{total}} f_{sig} +  \frac{(N_{total}-
  N_{sig})}{N_{total}} f_{back}
\end{equation}
where the free parameter $N_{sig}$ is the number of signal events and
$N_{total}$ is the total number of candidates in each momentum bin.
The function $f_{sig}$ is the Gaussian signal mass function:
\begin{equation}
  f_{sig}=\frac{1}{\sqrt{2\pi}s\sigma_{i}}e^{-\frac{1}{2}\left(
      \frac{M_{i}-M}{s \sigma_{i}} \right)^{2}}
\end{equation}
where $M_{i}$ is the candidate mass obtained from a kinematic fit of
the muon and kaon tracks.  The uncertainty $\sigma_{i}$ on the mass
is scaled by a free parameter $s$ in the unbinned maximum likelihood 
fit which is typically $\approx 1.2$. The parameter $M$ is the mean 
$B$ mass obtained by fitting Figure \ref{fig:BmassDist}.
The background mass function is linear:
\begin{equation}
  f_{back}= b \left( M_{i}-\frac{w}{2} \right)+\frac{1}{w}
\end{equation}
where $b$ is the slope of the background and $w$ is the
mass range in the fit ($5.15$ to $6.0$ GeV/$c^2$).  The region well above
the $B$ mass yields a better estimate of the slope of the background
since it is not affected by partially reconstructed $B$ decays.  The
likelihood function is minimized with respect to the parameters
$N_{sig}$, $s$ and $b$.  The fit yields $160 \pm 23$, $114 \pm 17$,
$62\pm 13$ and $71\pm 10$ events in the four transverse momentum bins.

\subsection{Acceptances and Trigger Efficiencies}

The acceptance is determined from a Monte Carlo simulation based on a
next-to-leading-order QCD calculation~\cite{QCDcalc} using the MRST
parton distribution functions~\cite{MRST}. The $b$-quark pole mass 
$m_b$ is taken to be $4.75$ \gevcc\bspace. The $b$ quarks are
produced in the rapidity range $|y_b| < 1.1$ with $p_T(b) > 5.5$
\gevc\bspace. The renormalization scale is $\mu = \mu_0 \equiv \sqrt{m_b^2 +
  p_T^2(b)}$, and the fragmentation scale is equal to the
renormalization scale. The fragmentation into \Bms is modeled using
the Peterson fragmentation function~\cite{pete} with the parameter
$\epsilon_P$ set to $0.006$~\cite{peterson}. This value was extracted 
in a fit to data collected at $e^+e^-$ colliders. Recent results from LEP and SLD
suggest that lower values of $\epsilon_P$ and other functions 
better describe the fragmentation of $b$ quarks into $B$ hadrons~\cite{SLDLEP}. 
Futhermore the 
assumption that a fragmentation function extracted from 
$e^+e^-$ data is an accurate description of $b$ fragmentation 
at a $p\bar p$ collider lacks a strong theoretical basis~\cite{Mangano}.  
However, the uncertainties due to these 
factors are expected 
to be smaller than the uncertainty on the renormalization scale.

Decays of Monte-Carlo-generated \Bms into the \jpsi and kaon final
states are performed using a modified version of the CLEO Monte Carlo
program~\cite{qq} which accounts for the expected 
\jpsi longitudinal polarization. 
Once the $B$ mesons are generated and decayed
into their final state, a simulation of the CDF detector is utilized.
A simulation of the trigger efficiency has also been included in the
acceptance calculation.  The events are then processed by the same
analysis code used on the data to determine the combined acceptance
and trigger efficiency for each momentum bin. The Run 1A and 1B results 
which incorporate  different trigger requirements are listed
in Table \ref{tbl:TrigAccept} together with the combined results.  
The uncertainties given are statistical only.
\begin{table}[htbp]
  \begin{center}
    \caption{The product of the trigger efficiency and the acceptance
             in the $p_{T}$ bins for Run 1A, Run 1B and the integrated
             luminosity-weighted average for Run 1.}
    \begin{tabular}{|c|c|c|c|} \hline 
      $p_{T}$ range &
      \multicolumn{3}{|c|}{ Trigger efficiency $\times$ acceptance (\%) }\\
      \hline
      (GeV/$ c $) & Run 1A & Run 1B & Run 1 \\
      \hline \hline
      6--9 &  2.01 $\pm$ 0.02 &  1.61 $\pm$ 0.02 &  1.70 $\pm$ 0.02 \\
      \hline
      9--12 &  5.29 $\pm$ 0.05 &  4.20 $\pm$ 0.04 &  4.44 $\pm$ 0.03 \\
      \hline
      12--15 &  8.36 $\pm$ 0.10 &  6.53 $\pm$ 0.09 &  6.93 $\pm$ 0.07 \\
      \hline
      15--25  & 11.96 $\pm$ 0.14 &  9.26 $\pm$ 0.12 &  9.86 $\pm$ 0.10 \\
      \hline 
    \end{tabular}\vspace{0.3cm}
    \label{tbl:TrigAccept}
  \end{center}
\end{table}

\subsection{Efficiencies of the Additional Selection Requirements}

The detector acceptance and trigger efficiencies described in the
previous section did not account for all of the criteria for selecting a
$B$ candidate. The efficiencies of the additional selection
requirements are discussed in this section. Most of these efficiencies
are determined using large CDF data samples.

There are two components that comprise the tracking efficiencies. The
first part is the efficiency of the tracking in the Level 3 trigger
system which is determined using an inclusive
single muon data set. The efficiency is measured to be $(97 \pm 2)\%$ for Run~1B.
During Run 1A, a portion of the data-taking suffered from the start
time of each event being incorrectly determined.
The result was an inefficiency in
reconstruction at Level 3 which was determined to be $\sim
4\%$~\cite{l3effi} averaged over all of Run 1A. The Level 3 Run 1A
efficiency is $(93 \pm 2)\%$.

Once an event has been accepted at Level 3, one
must account for the offline CTC track reconstruction which may
improve the muon track quality or find new tracks that are missed at Level 3.
It is also necessary to correct for the track finding efficiency for the
kaon track since it is not required in the Level 3 trigger.  
A detailed study~\cite{thesis} of the CTC track
reconstruction efficiencies was conducted.  
To measure the efficiency, we simulate single kaon tracks with the CDF
Monte Carlo.  We then combine the generated CTC hits for such a kaon
with the hits in an event with an identified displaced $J/\psi$ from the
CDF data sample.  Hits in the CTC are characterized by a leading edge
and a time-over-threshold.  Where a real and simulated hit overlap, the
hits are combined.  Thus the leading edges used in the track
reconstruction may be obscured for the simulated kaon as they would be
for real particles.  We then run the full track reconstruction program
on the modified event and search for a track corresponding to the
embedded kaon.  
We find the efficiency of the track reconstruction to be 
$(99.6^{+0.4}_{-0.9})\%$ for particles with
$p_T$ $>$ 0.8~GeV$/c$ that traversed all layers of the CTC, independent of
instantaneous luminosity. The Run 1A single track reconstruction 
efficiency of $(98.5\pm 1.4)\%$ is taken from Ref.~\cite{CDFold}.
 
The muon segment reconstruction efficiency is found to be $(98.0\pm 1.0)\%$
resulting in a combined efficiency of $(96.0\pm 1.4)\%$.
The efficiency of requiring both muons from the \jpsi to have a muon 
chamber track segment that  
matches a track reconstructed in the CTC is found to be $(98.7 \pm 0.2)\%$. 
The efficiency of this cut is determined from a sample of \jpsi candidate
events containing muons that were required to pass less stringent
matching requirements at Level 3.

The fraction of events in which both muons from the \jpsi have been
reconstructed in the SVX is measured using a large \jpsi data set.
This fraction is  
$(52.4 \pm 0.6)\%$ for Run 1A and $(56.3 \pm 0.2)\%$ for Run 1B.  
The fraction for Run 1B is larger than Run 1A
because the inner layer of the SVX detector was moved closer to the
beamline, eliminating a small separation between 
silicon wafers in the first layer present in Run 1A. 

The efficiency to reconstruct a $B$
meson with a proper decay length $ct$
greater than 100 \mm is determined using Monte Carlo simulations.  The
$ct$ resolution is measured in the \jpsi data set by fitting the
proper lifetime of events in the sidebands of the $B$ candidate mass
distribution with a Gaussian function for the prompt component and an
exponential function for the long-lived component. The
lifetimes of the Monte Carlo generated events are then smeared using
the resolution measured in each \pt range.  The efficiency showed no
significant variation with the $B$ transverse momentum even though
the proper $ct$ resolution was degraded by a factor of 2 from the 
lowest to the highest \pt
bin. The efficiency of $(78.4 \pm 0.5)\%$ is the mean of the values
measured in each $p_T$ bin.

The reconstruction efficiencies are summarized in Table
\ref{tbl:ReconEff}.  For the $B$ candidates decaying to particles
completely contained within the detector acceptance, the
reconstruction efficiency is $(36.4\pm 1.0)\%$.%

\begin{table}[htbp]
  \begin{center}
    \begin{tabular}{|c|c|c|} \hline 
    Source        & \multicolumn{2}{|c|}{Efficiency in \%} \\
                  &        Run1A        &      Run1B          \\ \hline \hline
    CTC tracking  & $(98.5\pm 1.4)^{3}$ &      $(99.6^{+0.4}_{-0.9})^{3}$\\ 
                  &  $=95.6\pm 2.4$   &  $=98.8^{+0.7}_{-1.5}$   \\ \hline 
L3 $\mu^+\mu^-$ tracking &   $93\pm 2$       &  $97\pm 2 $       \\ \hline
CTC-$\mu$ linking & \multicolumn{2}{|c|}{$(99.8\pm 0.2)^{2}$} \\
                  & \multicolumn{2}{|c|}{ $=99.6\pm 0.3 $}  \\ \hline
     Muon chamber & \multicolumn{2}{|c|}{$(98.0\pm 1.0)^{2}$} \\
     efficiency   & \multicolumn{2}{|c|}{$=96.0\pm 1.4$}    \\ \hline
$\mu^+\mu^-$ matching cut& \multicolumn{2}{|c|}{$98.7\pm 0.2$} \\ \hline
     $Z$ vertex cut & $95.3\pm 1.1$  &  $93.7\pm 1.1$  \\ \hline
    SVX fraction  &   $52.4\pm 0.6$ &  $56.3\pm 0.2$    \\ \hline 
      $ct>$ 100 $\mu$m   & \multicolumn{2}{|c|}{$78.4\pm 0.5$}  \\ \hline\hline 
      Total       & \multicolumn{2}{|c|}{$36.4\pm 1.2$}     \\ \hline
    \end{tabular}
  \end{center}
  \caption{Summary of reconstruction efficiencies for the $B$
    meson.  The efficiencies that are not common between 1A and 1B are
    averaged and weighted by integrated luminosity.}
  \label{tbl:ReconEff}
\end{table}

\subsection{Luminosity Determination}
%
%

At CDF the luminosity is measured using two telescopes 
of beam-beam counters to an accuracy of about 4\%.   
We studied the quality of the integrated luminosity calculation in the
inclusive $J/\psi\rightarrow\mu^+\mu^-$ sample.  After correcting for
the time-dependent trigger efficiency, we found that in Run 1B the measured
$J/\psi$ cross section $\sigma_\psi$ fell linearly as a function of
instantaneous luminosity $\cal L$.  However, for any narrow range of
$\cal L$, $\sigma_\psi$ was constant as a function of time. 
Since the minimum luminosity of the data sample is
$4\times10^{30}\,{\rm cm}^{-2}{\rm s}^{-1}$,
we have considered two possible extrapolations of $\sigma_\psi$ as 
a function of $\cal L$ to ${\cal L}=0$ to calculate a corrected 
integrated luminosity. The first extrapolation is performed 
assuming that the linear dependence is valid 
below ${\cal L}<4\times10^{30}$ and that:
\begin{equation}
  \smallint{\cal L}^\prime dt= \int {\cal
  L}(t){\sigma_\psi(0)\over\sigma_\psi({\cal L})}dt
\end{equation}
We also perform the extrapolation assuming that no correction 
is needed below ${\cal L}<4\times10^{30}$. The luminosity 
correction is taken to be the average of the two extrapolations and
we assign a systematic uncertainty that covers the range between 
the two hypotheses. The 
correction to the integrated luminosity for Run 1B is 
\begin{equation}
R_{\cal L}\equiv\smallint{\cal L}dt/\smallint{\cal L}^\prime dt= 0.88\pm0.04.
\end{equation}

\subsection{Systematic Uncertainties}

We divide the systematic uncertainties in the measurement of the $B^+$
meson production cross section into two classes: \pt dependent 
uncertainties $(syst_{\pt})$ that change from one \pt bin to the next
and fully correlated uncertainties $(syst_{fc})$ that are independent of
\pt\bspace.

\subsubsection{\pt dependent systematic uncertainties}

The \pt dependent systematic uncertainties include variations of the
production and decay kinematics that would affect the determination of
the acceptance.  We have considered effects due to the model used to
generate the $b$ quark spectrum and uncertainties in our knowledge of
the trigger efficiency.

The model used to generate the $b$ quarks is based on a QCD
calculation at next-to-leading order.  Large uncertainties in the
calculation are due to unknown higher-order effects. These effects are
quantified by estimating the scale dependence when the renormalization
and factorization scales are varied by a factor of 2 above and below
their central value of $\mu=\mu_0=\sqrt{p_T^2+m_b^2}$.  The Peterson
fragmentation parameter is varied by $\pm 0.002$ around its central
value of $\epsilon_P=0.006$.  In each case the uncertainty on the
acceptance is taken to be the difference between the acceptance found
with the central value and the value found when each variable is
varied by the indicated amounts.  The dependence of acceptance on the parton
density parametrization and the $b$ quark mass are much smaller and
are not included in the systematic uncertainty. In addition, the 
parameters of the trigger simulation are varied 
by $\pm 1 \sigma$.  The
total \pt dependent uncertainty is given by the sum in quadrature of the
\pt dependent systematic uncertainties summarized in Table III.

\subsubsection{Correlated systematic uncertainties}

The correlated systematic uncertainties include uncertainties
that are independent of the $B$ meson $p_T$ spectrum.
The largest of these uncertainties is due to limited knowledge 
of the $B^+ \rightarrow J/\psi K^+ $ branching ratio~\cite{PDG} which 
yields a systematic uncertainty of about 10\%.
Other sources of correlated uncertainties are due 
to the uncertainty on the total reconstruction
efficiency shown in Table II and knowledge of the integrated luminosity 
collected at CDF during 
Run 1. There is an additional systematic uncertainty  
associated with 
the reconstruction of kaons that decay inside the CTC volume.
A simulation shows that
about $ 8\% $ of the kaons decay in flight, of which half are successfully
reconstructed~\cite{CDFold}. We assign the full value of the correction as
an uncertainty for the kaon acceptance of $(96 \pm 4)$\%. This
assumes that such tracks are modeled realistically in the simulation.
The total correlated uncertainty of $12.7\%$ is given by the sum in quadrature 
of the fully correlated systematic uncertainties summarized 
in Table IV.
\begin{table}[htbp]
  \begin{center}
  \caption{Summary of \pt dependent systematic uncertainties.}
    \begin{tabular}{|c|c|c|c|c|} \hline
      Source & \multicolumn{4}{|c|}{Fractional uncertainty in each \pt bin} \\ \hline
                     $p_T$ range (GeV/$c$) 
& ~~~6--9~~~  & ~~~9--12~~~ &  ~~~12--15~~~  & ~~~15--25~~~  \\ \hline \hline
      QCD renormalization uncertainty &  1.6\% & 1.5\% & 1.7\% & 1.5\% \\ \hline
      Peterson parameter uncertainty  &  0.7\% & 1.6\% & 1.0\% & 1.7\% \\ \hline
      Trigger efficiency uncertainty  &  3.1\% & 2.7\% & 2.1\% & 1.7\% \\ \hline \hline
      \pt dependent total (syst$_{\pt}$) &  3.6\% & 3.5\% & 2.9\% & 2.8\% \\ \hline 
    \end{tabular}
  \end{center}
  \label{tbl:UncorrSys}
\end{table}

\begin{table}[htbp]
  \begin{center}
    \caption{Summary of fully correlated systematic uncertainties.}
    \begin{tabular}{|c|c|} \hline
      Source                              & Fractional uncertainty        \\ \hline \hline
      Reconstruction efficiency           &  $\pm 2.7\%$ \\ \hline
      Luminosity uncertainty              &  $\pm 4.1\%$ \\ \hline
      Luminosity correction               &  $\pm 4.5\%$ \\ \hline 
      Branching ratio uncertainty         &  $\pm 10.2\%$ \\ \hline
      Kaon decay-in-flight uncertainty    &  $\pm 4.0\%$ \\ \hline  \hline
      Fully correlated total (syst$_{fc}$) & $\pm 12.8\%$ \\ \hline  
    \end{tabular}
  \end{center}
  \label{tbl:CorrSys}
\end{table}

\subsection{Results}

The differential cross section $ d\sigma /dp_{T} $ is calculated using
the following equation:
\begin{equation}
  \frac{d\sigma (B^{+})}{dp_{T}}= \frac{N_{sig}/2}{\Delta p_{T} \cdot
  {\cal L}^{\prime} \cdot A \cdot \epsilon \cdot
  {\mathcal{B}}}
\end{equation}
where $N_{sig}$ is the number of charged $B$ mesons determined from
the likelihood fit of the mass distribution in each \pt range. The
factor of 1/2 is included because both $B^{+}$ and $ B^{-} $ mesons
are detected while we report the cross section for $B^+$ mesons  
assuming charge invariance in the production process.
The width of the \pt bin is $\Delta p_{T}$ and $\mathcal{L}^{\prime} $ is the
corrected integrated luminosity of the sample. The geometric and kinematic
acceptance $A$ is determined from the Monte Carlo simulation and includes
the kinematic and trigger efficiencies. The efficiency $\epsilon $ is
the additional reconstruction efficiency not included in the
simulation.  The product of branching ratios $\mathcal{B}$ is
determined using the the world-average~\cite{PDG} branching fractions:

\begin{equation}
  BR(B^{\pm }\to J/\psi ~K^{\pm })=(10.0 \pm 1.0)\times 10^{-4}
\end{equation}
\begin{equation}
  BR(J/\psi \to \mu ^{+}\mu ^{-})=(5.88\pm 0.10)\times 10^{-2}
\end{equation}

Table V lists the differential cross section as a function of $ p_{T}$. 
The three uncertainties quoted on the cross section are statistical
(stat), \pt dependent systematic (syst$_{\pt}$), and
fully correlated systematic (syst$_{fc}$),
respectively.

\begin{table}[htbp]
  \begin{center}
  \caption{\protect$ B^{+}\protect $ meson differential cross section
  from the Run
    1 data.}
    \begin{tabular}{|c|c|c|c|} \hline 
      $\left\langle p_{T}\right\rangle $ & Events & Acceptance & Cross section \\
      (\gevc) & & (\%) & (nb/[GeV/$c$]) \\ \hline \hline
      7.34  & 160 $\pm$ 23 &  1.70 $\pm$ 0.02 & 815  $\pm$ 117(stat)
            $\pm$ 31(syst$_{\pt}$) $\pm$ 104(syst$_{fc}$)\\ \hline
      10.35 & 114 $\pm$ 17 &  4.44 $\pm$ 0.03 & 222  $\pm$  33(stat)
            $\pm$ 8(syst$_{\pt}$) $\pm$  28(syst$_{fc}$)\\ \hline
      13.36 &  62 $\pm$ 13 &  6.93 $\pm$ 0.07 & 77.5 $\pm$  16.2(stat)
            $\pm$ 2.4(syst$_{\pt}$) $\pm$ 9.9(syst$_{fc}$)\\ \hline
      18.87 &  71 $\pm$ 10 &  9.86 $\pm$ 0.10 & 18.7 $\pm$ 2.6(stat)
            $\pm$ 0.6(syst$_{\pt}$) $\pm$  2.4(syst$_{fc}$)\\ \hline
    \end{tabular}
  \end{center}
  \label{tbl:DiffXsec}
\end{table}

\begin{figure}[htbp]
  \centering
  \includegraphics[width=10cm]{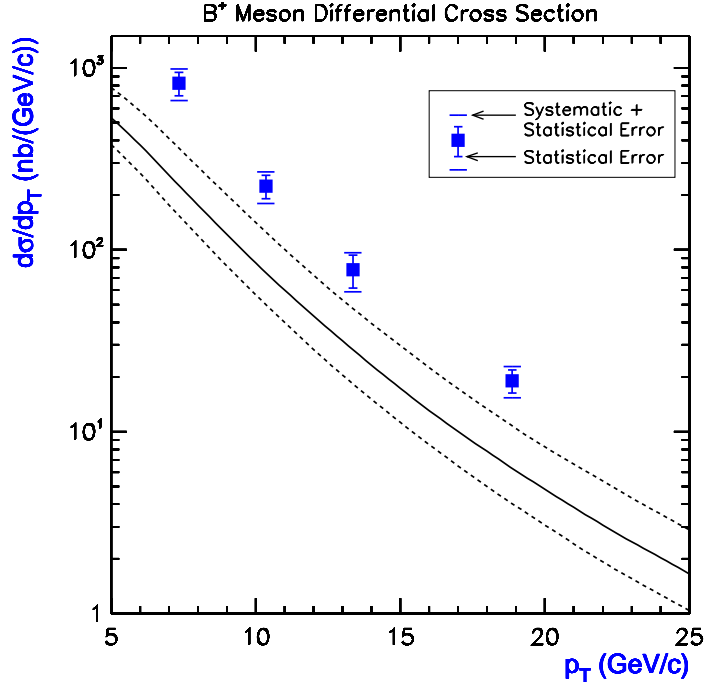}
  \caption{$B^+$ meson differential cross measurements compared to the
theoretical prediction. The solid curve is the theoretical prediction
for $m_b = 4.75$ GeV/$c^2$, $\mu_0 =\sqrt{m_b^2 + p_T^2}$, 
$\epsilon_P = 0.006$ and $f_u = 0.375$.
The dashed lines illustrate the changes in the theory once these parameters 
are varied as explained in the text.} 
  \label{fig:DiffXsecTheory}
\end{figure}

Figure \ref{fig:DiffXsecTheory} shows the measured differential cross
section at the mean \pt of each bin compared to the next-to-leading-order
QCD~\cite{QCDcalc} calculation using the MRST parton density
functions~\cite{MRST}. The experimental points are plotted at $\langle p_T \rangle $
which is the value of $p_T$ for which the theoretical differential
cross section~\cite{MRST} equals the mean cross section in each
momentum range
\begin{equation}
\left.  \frac {d\sigma}{d p_T} \right|_{\langle p_T \rangle}=
\frac{1}{\Delta p_T} \int^{\Delta p_T}{ \frac{d\sigma}{d p_T} }d p_T
\end{equation}

The dashed lines in Figure \ref{fig:DiffXsecTheory} indicate the change in the
theoretical predictions as the $b$ quark mass is varied between $4.5$
and $5.0$ \gevcc\bspace, the renormalization scale is varied between
$\mu_0/2$ and $2\mu_0$, and the Peterson fragmentation parameter is
varied between 0.004 and 0.008. The solid curve is for the central
values of these parameters: $m_b = 4.75$ GeV/$c^2$, $\mu_0 =
\sqrt{m_b^2 + p_T^2}$, and $\epsilon_P = 0.006$. The fraction of $\bar
b$ quarks that fragment into $B^+$ is $f_u = 0.375 \pm 0.023$~\cite{CDFfrag}. 
This fraction is varied between 0.352 and 0.398.

\begin{figure}[htbp]
  \centering
  \includegraphics[width=10cm]{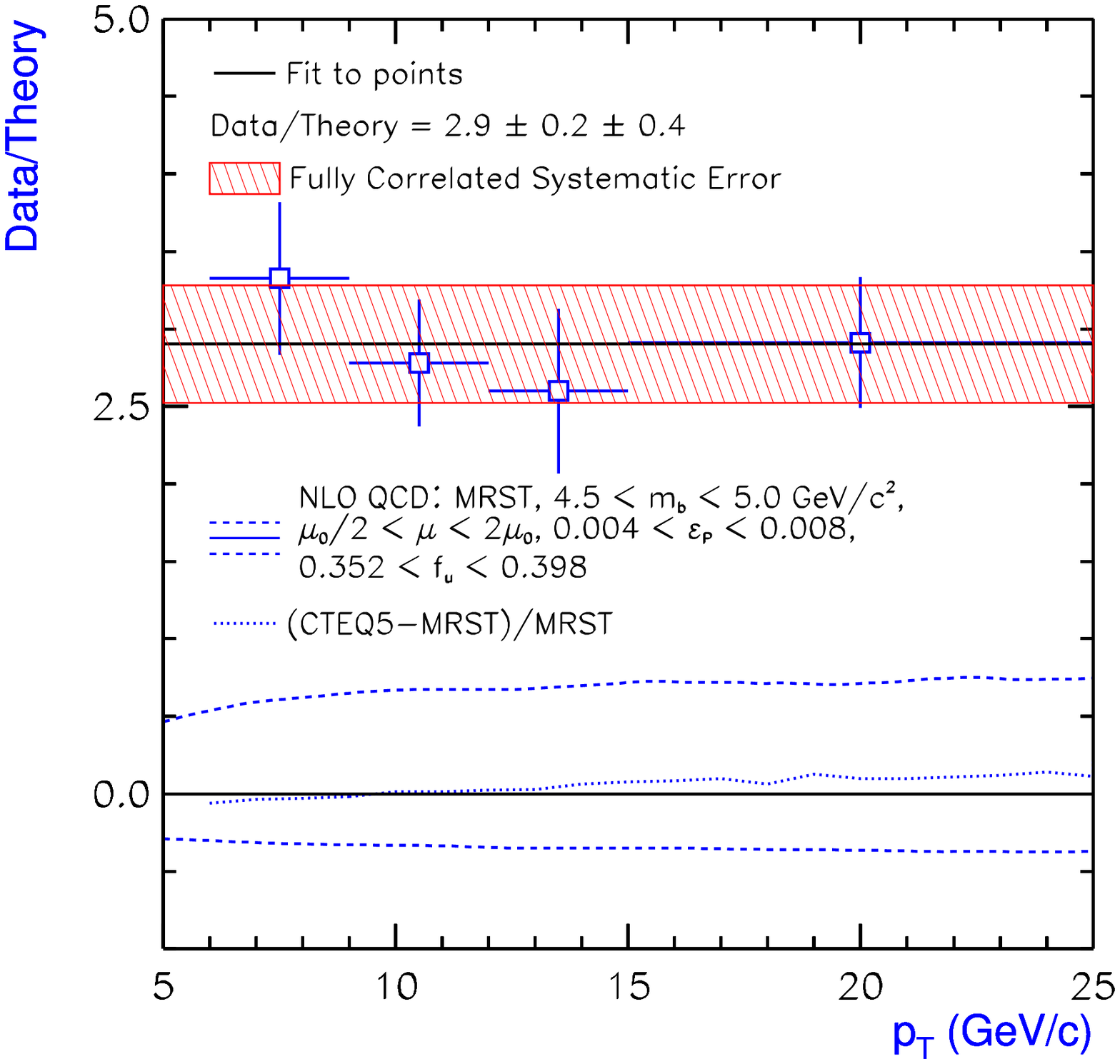}
  \caption{Plot of data/theory as a shape comparison with 
    the NLO QCD differential cross section calculations.}
  \label{fig:Ratio}
\end{figure}

The comparison between data and theory for $d\sigma/dp_T$ is aided by 
plotting the ratio of data/theory on a linear scale, as shown in Figure
\ref{fig:Ratio}.  The level of agreement between the data and the
theoretical prediction is determined by fitting a line through the
four ratio points.  The fit yields a scale factor for data/theory of
$2.9 \pm 0.2~{\rm (stat \oplus syst_{\pt})} \pm 0.4~{\rm (syst_{fc}) } $ with
a confidence level of 72\%.  
The first uncertainty on the scale factor is the uncertainty returned by the fit 
to the ratio points whose uncertainties were determined by summing 
the statistical and the \pt dependent systematic uncertainties in quadrature. 
The second uncertainty is the fully correlated systematic uncertainty.
The hatched band shows the magnitude of the fully correlated uncertainty
which arises mainly due to the poor knowledge of the $B^+ \rightarrow
J/\psi K^+$ branching franction. Also shown is a comparison between
the shape of the QCD predictions obtained using a different set of
parton distribution functions determined by the CTEQ
collaboration~\cite{CTEQ}.  The effect of changing the parton
distribution functions is negligible in comparison with the variation
associated with uncertainties in the $b$ quark mass, the fragmentation
parameter and the renormalization scale shown by the dashed curves. 

\section{The total cross section \label{s:total}}

\begin{figure}[htbp]
\centering
\includegraphics[width=10cm]{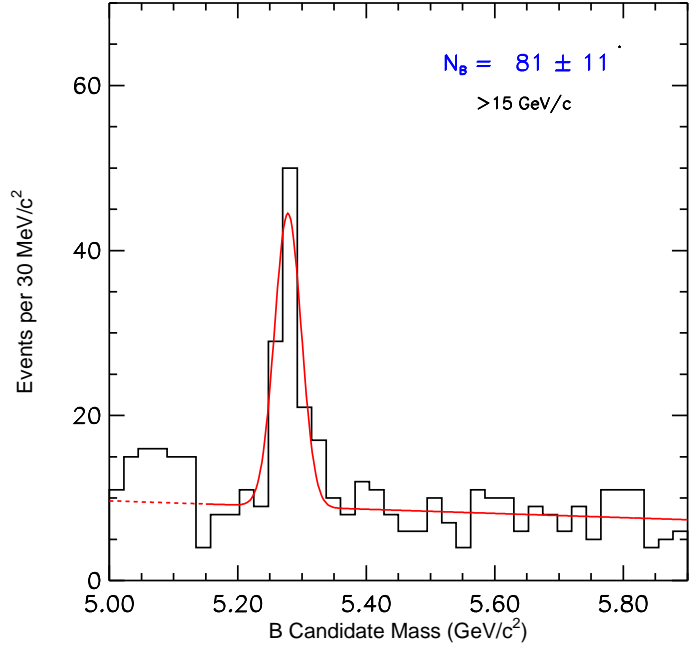}
\caption{$B^{\pm}$ candidate mass distribution for $p_T(B)$ $>$ 15 GeV/$c$.
The curve is a binned fit to a Gaussian distribution plus linear 
background and is for illustration only.} 
\label{fig:lastbin}
\end{figure}

The total cross section is obtained by using a method similar to the
one used for the determination of the differential cross section.
However, the last tranverse momentum bin, 15--25 GeV/$c$, is
replaced with the invariant mass distribution for $B^{\pm }$ candidate
events with $p_T>$ 15 \gevc shown in Fig. \ref{fig:lastbin}.  With 81
$\pm$ 11 candidates and an acceptance of (10.19 $\pm$ 0.16)\%, the
integrated cross section for $p_T > 15$ GeV/$c$ is 207 $\pm$ 28(stat) $\pm$
5(syst$_{\pt}$) $\pm$ 26(syst$_{fc}$)~nb.  The
integrated cross section for  $B$ transverse momentum  
$p_T > 6.0$ GeV/$c$  and $|y|<1.0$ is given by:
\begin{equation}
  \sigma (B^{+})= \sum_{i=1}^{4} \frac{N_i/2}{{\cal L}^{\prime} \cdot A_i \cdot
  \epsilon \cdot {\mathcal{B}}} 
\end{equation} 
where $N_i$ is the number of charged $B$ candidate events in each
momentum bin, $A_i$ is the acceptance and $\epsilon$ is reconstruction
efficiency.  The total cross section is:
\begin{equation}
  \sigma_{B}(p_{T} > 6.0\, {\rm GeV/}c,\, |y| <
  1.0) = 3.6 \pm 0.4~(\rm{stat} \oplus \rm{syst_{\pt}}) \pm 0.4~ ({\rm
  syst_{fc}})\, \mu {\rm b}.
\end{equation}
where the first uncertainty is the sum in quadrature of the statistical and
\pt dependent systematic uncertainty, and the second uncertainty is the fully
correlated systematic uncertainty.

\section{Summary\label{s:summary}}

The exclusive decay \bdecay has been used to measure the production
cross section of the $B^+$ meson from data collected by the CDF
detector.  A sample size of $387 \pm 32$ events is obtained from
$\int{{\mathcal L}} dt =$ \intlumx of 1.8 TeV $p \bar p$
collisions produced by the Fermilab Tevatron collider. 

The measured total $B^+$ production cross section for $p_T(B) >
6.0$ \gevc and $|y|<1.0$ is
\begin{equation}
  \sigma_{B}(p_{T} > 6.0\, {\rm GeV/}c,\, |y|<
  1.0)=3.6 \pm 0.6 ({\rm stat} \oplus {\rm syst)} \, \mu {\rm b} 
\end{equation}
where the uncertainty is the sum in quadrature of the
statistical and both correlated and \pt dependent systematic uncertainties.
The differential cross section is
measured to be $2.9 \pm 0.2 ~({\rm stat \oplus syst_{\pt}}) \pm 0.4 ~({\rm
  syst_{fc}})  $ times higher than the NLO QCD predictions with
agreement in shape.  
The first uncertainty is the sum in quadrature of the
statistical and \pt dependent systematic uncertainty and the second is the 
correlated systematic uncertainty.
The new measurement of the $B^+$
differential cross section confirms that the absolute rate is
larger than the limits of that predicted by typical variations in the
theoretical parameters. 

These measurements supersede those of reference~\cite{CDFold}.

\clearpage



\begin{acknowledgments}
We thank the Fermilab staff and the technical staffs of the
participating institutions for their vital contributions.  This work
was supported by the U.S. Department of Energy and National Science
Foundation; the Italian Istituto Nazionale di Fisica Nucleare; the
Ministry of Education, Science, Sports and Culture of Japan; the
Natural Sciences and Engineering Research Council of Canada; the
National Science Council of the Republic of China; the Swiss National
Science Foundation; the A.P. Sloan Foundation; the Bundesministerium
fuer Bildung und Forschung, Germany; and the Korea Science and
Engineering Foundation.

\end{acknowledgments}

\bibliography{basename of .bib file}

\begin{thebibliography}{999}

\bibitem{QCDcalc}
Nason {\it et al.}, Nucl. Phys. {\bf B327}, 49 (1989), erratum
{\it ibid.} {\bf B335}, 260 (1990);
Beeneker {\it et al.}, Nucl. Phys. {\bf B351}, 505 (1991).


\bibitem{Mangano}
S.~Frixione, M.~Mangano, P.~Nason, and G.~Ridolfi, 
Heavy Flavors II, eds. A.J. Buras and M. Lindner, Advanced
Series on direction in High Energy Physics, World Scientific Publishing Co.,
Singapore, 1997.

\bibitem{ua1}
UA1 Collaboration, C. Albajar {\it et al.}, 
Phys. Lett. B {\bf 186}, 237 (1987); {\bf 256}, 121 (1991).

\bibitem{tevatron}
CDF Collaboration, F. Abe {\it et al.}, Phys. Rev. 
Lett. {\bf 71}, 500 (1993); {\it ibid.} {\bf 79}, 572 (1997)
Phys. Rev., Lett. {\bf 75}, 1451 (1995); D0 Collaboration,
B. Abbot {\it et al.}, Phys. Lett. B {\bf 487}, 264 
(2000); D0 Collaboration,
B. Abbot {\it et al.}, hep-ex/0008021.
 
%
\bibitem{Berger}
E. L. Berger, B. W. Harris, D. E. Kaplan, Z. Sullivan,
T. M. P. Tait, C. E. M. Wagner, Phys.Rev. D {\bf 63}, 115001 (2001). 



\bibitem{CDFold}

CDF Collaboration, F. Abe {\it et al.}, 
Phys. Rev. Lett. {\bf 75}, 1451 (1995).


\bibitem{Cleo} M. S. Alam {\it et al.}, Phys. 
Rev. D {\bf 50}, 43 (1994).

\bibitem{CDFdet}
CDF Collaboration, F. Abe  {\it et al.}, Nucl. Instrum. Methods
Phys. Res., Sect. {\bf A271}, 388 (1988).

\bibitem{CDFsvx}
S. Tkaczyk  {\it et al.}, Nucl. Instrum. Methods
Phys. Res., Sect. {\bf A342}, 240 (1994); D. Amidei  {\it et al.}, Nucl. Instrum. Methods
Phys. Res., Sect. {\bf A360}, 137 (1995).

\bibitem{resol}
D. Amidei  {\it et al.}, Nucl. Instrum. Methods
Phys. Res., Sect. {\bf A350}, 73 (1994).


\bibitem{CFT}

G. Foster  {\it et al.}, Nucl. Instrum. Methods
Phys. Res., Sect.{\bf A269}, 93 (1988).


\bibitem{PDG}
Particle Data group, C. Caso {\it et al.}, Eur. Phys. J. C {\bf 3}, 1 (1998). 


\bibitem{CDFlife}
CDF Collaboration, F. Abe {\it et al.}, 
Phys. Rev. D {\bf 57}, 5382 (1998);
Phys. Rev. D {\bf 59}, 32004 (1999);
Phys. Rev. D {\bf 58}, 92002 (1998).


\bibitem{MRST}
A. Martin, W. Stirling and R. Roberts, Phys. Lett. B {\bf 306}, 145 (1993);
Phys. Lett. B {\bf 443}, 301 (1998); Eur. Phys. J. C {\bf 4 }, 463 (1998).


\bibitem{pete}
C. Peterson {\it et al.}, Phys. Rev. D  {\bf 27}, 105 (1983).

\bibitem{peterson}
J. Chrin, Z. Phys. C {\bf 36}, 163 (1987).

\bibitem{SLDLEP}
OPAL Collaboration, G. Alexander {\it et al.}, Phys. Lett. 
B {\bf 364}, 93 (1995);
SLD Collaboration, K. Abe {\it et al.}, Phys. Rev. Lett. 
{\bf 84}, 4300 (2000);
ALEPH Collaboration, A. Heister {\it et al.}, Phys. Lett. 
B {\bf 512}, 30 (2001).


\bibitem{qq}

P. Avery, K. Read and G. Trahern, ``QQ: A Monte Carlo Generator'', CLEO 
Internal Software Note
CSN-212, Cornell University, 1985.

\bibitem{l3effi}
CDF Collaboration, F. Abe {\it et al.}, Phys. Rev. Lett. {\bf 75}, 4358 (1995).


\bibitem{thesis}
T. Keaffaber, ``Measurement of the $B^+$ Meson Cross section in Proton-Antiproton Collisions
at 1.8 TeV Using the Fully Reconstructed Decay 
$B^+ \rightarrow J/\psi K^+$, Purdue University Dissertation, 2000.

\bibitem{CDFmumu}
CDF Collaboration, F. Abe {\it et al.}, 
Phys. Rev. D {\bf 57}, R3811 (1998).


\bibitem{CDFfrag}
CDF Collaboration, T. Affolder {\it et al.}, Phys. Rev. D {\bf 60}
92005, (1999); CDF Collaboration, T. Affolder {\it et al.}, Phys. Rev. Lett. {\bf 84}
1663, (2000).

\bibitem{CTEQ}
H. L. Lai {\it et. al}, ``Global QCD Analysis of Parton Structure of the 
Nucleon: CTEQ5 Parton Distributions'', hep-ph/9903282, (1999).

\end{thebibliography}

\end{document}